# A magnetic diverter for charged particle background rejection in the SIMBOL-X telescope

D. Spiga[a*], V. Fioretti[b], A. Bulgarelli[b], E. Dell'Orto[a,d], L. Foschini[b], G. Malaguti[b], G. Pareschi[a], G. Tagliaferri[a], A. Tiengo[c]

[a]INAF/ Osservatorio Astronomico di Brera, Via E. Bianchi 46, 23807 Merate (LC), Italy
[b]INAF/ Istituto di Astrofisica Spaziale e Fisica Cosmica, Via Gobetti 101, 40129 Bologna, Italy
[c]INAF/ Istituto di Astrofisica Spaziale e Fisica Cosmica, via Bassini 15, 20133 Milano, Italy
[d]Università degli Studi dell'Insubria, Via Ravasi 2, 21100 Varese, Italy

**ABSTRACT**

Minimization of charged particle background in X-ray telescopes is a well known issue. Charged particles (chiefly protons and electrons) naturally present in the cosmic environment constitute an important background source when they collide with the X-ray detector. Even worse, a serious degradation of spectroscopic performances of the X-ray detector was observed in Chandra and Newton-XMM, caused by soft protons with kinetic energies ranging between 100 keV and some MeV being collected by the grazing-incidence mirrors and funneled to the detector. For a focusing telescope like SIMBOL-X, the exposure of the soft X-ray detector to the proton flux can increase significantly the instrumental background, with a consequent loss of sensitivity. In the worst case, it can also seriously compromise the detector duration. A well-known countermeasure that can be adopted is the implementation of a properly-designed magnetic diverter, that should prevent high-energy particles from reaching the focal plane instruments of SIMBOL-X. Although Newton-XMM and Swift-XRT are equipped with magnetic diverters for electrons, the magnetic fields used are insufficient to effectively act on protons. In this paper, we simulate the behavior of a magnetic diverter for SIMBOL-X, consisting of commercially-available permanent magnets. The effects of SIMBOL-X optics is simulated through GEANT4 libraries, whereas the effect of the intense required magnetic fields is simulated along with specifically-written numerical codes in IDL.

## 1. INTRODUCTION

The need of minimizing the particle background in X-ray space telescopes is to date a well known issue. Charged particles (protons, electrons, and - to a lesser extent - heavier ions) naturally present in the cosmic environment constitute an important background source when they collide with the X-ray detector. Though the electron background was already foreseen for Chandra, Newton-XMM and Swift-XRT, the issue of "soft" protons background (with kinetic energies ranging between 50 keV and some MeV) came totally unexpected.

The problem was driven to attention when the Chandra X-ray Observatory was launched in 1999. Soon after the launch, an increase of the CTI (*Charge Transfer Inefficiency*) was observed on the ACIS[1,2] (Advanced CCD Imaging Spectrometer), The degradation was particularly evident at orbital crossing of the terrestrial magnetosphere[3,4], which is well known to be a reservoir of high-energy charged particles. At that time, the unwanted capability of the

---

[*] e-mail: daniele.spiga@brera.inaf.it, phone +39-039-5971027

X-ray mirror assembly to reflect and focus high-energy particles – in addition to X-rays – was underestimated, therefore the ACIS had been effectively shielded against particle background from all direction, but that of the mirrors. The range of soft protons in Silicon (0.92 μm for 100 keV protons[5]) is sufficient for them to traverse the top layers of the CCD and create charge traps in the sensitive region, resulting in an energy sensitivity degradation. In particular, for Chandra's ACIS detector the most dangerous protons seemed to be those with kinetic energies from 100 to 200 keV[6].

In order to preserve the CCD from a quick damage, the ACIS is moved out of the focal plane when the particle flux becomes too high. Moreover, passive protection of the X-ray detectors against proton irradiation was obtained by obstructing the detector field with the transmission grating when the ACIS is not in use[6]. Finally, the ACIS is not used when the satellite is at too small radial distances from Earth (12 terrestrial radii[3] in 2000).

The soft protons is clearly an issue also for Newton-XMM[7,8,9]. Indeed, as Newton-XMM exhibits a much higher grasp as Chandra, the proton irradiation on the MOS-CCD would have been much higher if it were not partially obstructed by the RGS gratings, without which the lifetime of MOS CCDs would have been limited to 1 day[10]! Even if Newton-XMM's CCDs are coated with protective layers (aluminum, polypropylene and tin for the EPIC-MOS camera onboard XMM), soft protons can traverse them and, after some energy loss, deposit their energy in the detector charge transfer region. They are detected there as if they were X-ray photons and heavily contribute to the background.

It is known now that soft protons originate from the Sun, being accelerated by magnetic field reconnection in Solar magnetosphere. These are trapped in the Earth magnetosphere and collected by focusing optics of X-ray telescopes when they traverse it. The proton flux is much higher when the telescope crosses the radiation belts (Fig. 1, left), with a maximum at satellite perigee. Moreover, it exhibits annual variations[3] due to the rotation of the geomagnetic tail and it is affected by solar flares activity (it may increase by a 10-fold factor with respect to the quiescent solar time[8]), therefore it is highly unstable and the resulting background is very difficult to subtract. In the radiation belts, the proton energy spectrum is a double power-law[6] with a slope change between 0.5 and 1 MeV. The proton flux out of radiation belts is mainly due to solar wind, but still relevant: the 100 keV proton flux at a 70000 km of altitude[11] is 3 $cm^{-2}$ $sec^{-1}$ $sr^{-1}$ $keV^{-1}$. The proton spectrum out of radiation belts exhibit a power-law signature with a power-law index close to 2 (see Fig. 1, right).

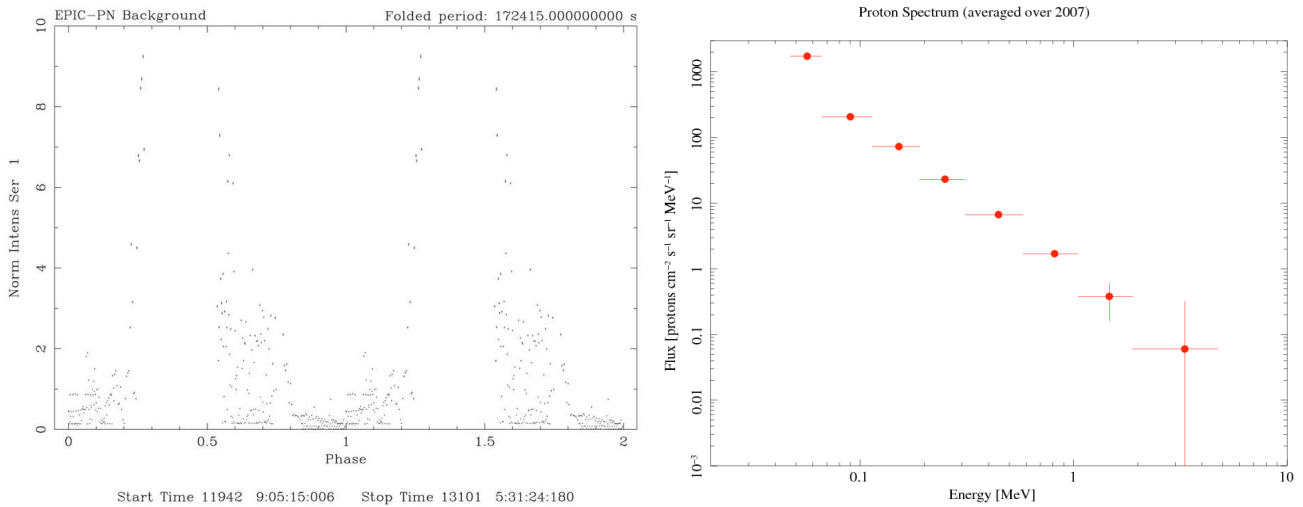

Fig. 1: (left) the measured background of EPIC-PN onboard Newton-XMM, as a function of the orbital phase. The increase of the measured background is apparent at the crossing of the radiation belts.
(right) The measured spectrum of the protons out of the Earth magnetosphere (Public data from the Advanced Composition Explorer (ACE) satellite obtained through the Coordinated Data Analysis Web of the NASA/GSFC).

In order to reduce the charged particle flux impinging onto the detectors, Newton-XMM and Swift-XRT are equipped with a *magnetic diverter for electrons*[12]. Magnetic fields are able to effectively shield the detectors from electrons, leaving unaffected the X-ray photon flux. Those electron diverters consist of permanent magnets fixed to the 12 spider spokes of the mirror module (to avoid X-ray vignetting), with magnetic dipoles oriented on the spider plane to produce a azimuthally-directed magnetic induction *B*. For the Swift telescope, the magnets were bars of dimension 15x5x70 mm, producing a magnetic field from 110 G in the mid-plane and 30 G at 2 cm from mid-plane. Similar values for the magnetic field are adopted for the electron diverter of Newton-XMM (25 to 150 G). These equipments are able to effectively shield electrons as they reduce the electron flux below 100 keV on the detector to less than $10^{-5} e^-/cm^2/sec$, much less than the background from other sources[12]. Their efficiency decreases for increasing electron kinetic energy. Nevertheless, they are unable to effectively deviate protons, that are 1836 times as massive as electrons.

Even if SIMBOL-X orbit will be mainly outside the radiation belts, the soft proton flux it will encounter is relevant also, therefore it will constitute a major background source also for the low-energy detector (LED) onboard the SIMBOL-X[13] X-ray telescope, with sensitivity up to 20 keV (the HED, the High-Energy-Detector, seems to be shaded by the LED against proton flux). As for Chandra, Newton-XMM, Swift-XRT, it is the optics that cause the particles to funnel toward the detector. In fact, grazing-incidence mirrors behave also like particle concentrators, even though the physical mechanism responsible for particle reflection has not been definitely established. Anyway, the particle flux can still be deviated by means of a proper magnetic field arrangement, able to generate a strong magnetic field in a region traversed by particles, even though the deflection efficiency also appears to depend on the angular distribution of protons and electrons at the optical module exit, which in turn depends on the actual reflection process of high-energy particles.

The feasibility of a proton diverter was already studied by Turner[14] for the XEUS telescope case. In this work we shall examine the feasibility of a magnetic diverter for the SIMBOL-X telescope. In Sect. 2 we briefly consider the different models proposed to describe how the protons are scattered off the mirrors surfaces. In Sect. 3 we describe a possible configuration for the magnetic diverter based on permanent magnets, while in Sect. 4 the magnetic field in the space surrounding the diverter is computed and we deal with some Monte-Carlo simulations to assess the effectiveness of the magnetic field configuration. The results and the open issues are briefly summarized in Sect. 5.

## 2. A REVIEW OF PROTON REFLECTION THEORIES

Protons are expected to emerge from optics more tightly focused than electrons[14]. This is a factor that eases the magnetic deflection of protons, since it also reduces the likelihood for an unfocused particle to be deflected by the magnetic field into the detector area. Even so, the effectiveness of the magnetic diverter for SIMBOL-X depends critically on the proton reflectivity and the angular and energetic distribution of the protons as reflected by the mirrors surfaces. Unfortunately, the physical mechanisms that might cause grazing incidence X-ray mirrors to behave also like particle concentrators is not cleared yet, thus we are not able to give a prediction for the angular spread introduced by the proton scattering off the SIMBOL-X mirrors. We list in the following paragraphs the physical models proposed in the last years to interpret the proton reflection from X-ray mirrors.

### 2.1. *Rutherford scattering* of protons

The multiple Rutherford scattering of protons in the X-ray reflective layer of the mirrors was suggested as the first plausible process able to scatter protons off X-ray mirrors. Protons enter the X-ray reflective coating and penetrate the electric field of nuclei (e.g. of Gold), where they get scattered. The deflection angle may be enhanced by repeated subsequent scattering processes. Due to the random nature of the process, we expect that the proton beam emerges from the reflection with a considerable angular spread. Moreover, in the deflection the protons transfer a recoil momentum to the reflective layer atoms, thus they dissipate a variable amount of kinetic energy: the energy transfer depends on the proton energy, the deflection angle and the number of scatterings. Therefore, we expect that a mono-energetic proton beam will emerge from the optic with a broader energy spectrum.

The multiple Rutherford scattering was implemented in the GEANT4 libraries in order to simulate the effect of proton scattering off the XMM mirrors. Application of GEANT4 routines to the XMM optics[11] returned a detailed description of the proton irradiation of EPIC-MOS and RGS detectors. We performed a preliminary simulation by running GEANT4 with a simplified SIMBOL-X mirrors structure, assuming a graded Pt/C multilayer as X-ray reflective coating. The simulation assumed 500000 protons of 500 keV kinetic energy to enter the optical module up to a 5 deg off-axis. The simulation results are displayed in Fig. 2: the angular distribution at the optic exit, shown on the left side, approximated a Gaussian distribution with a $\sigma$ = 2.8 deg. The energy distribution, as expected (Fig. 2, right side) exhibits a tail at lower kinetic energies, revealing an important inelasticity of the multiple scattering process.

The application of the multiple scattering approach was criticized by Aschenbach[10] because the potential needed to deflect a 100 keV proton by 0.5 deg would be found only inside the N shell of a Gold atom, that occupies only a small part of the layer volume. Moreover, comparison of experimental results obtained on XMM mirror samples[15] sorted out a doubtful accord with GEANT4 predictions.

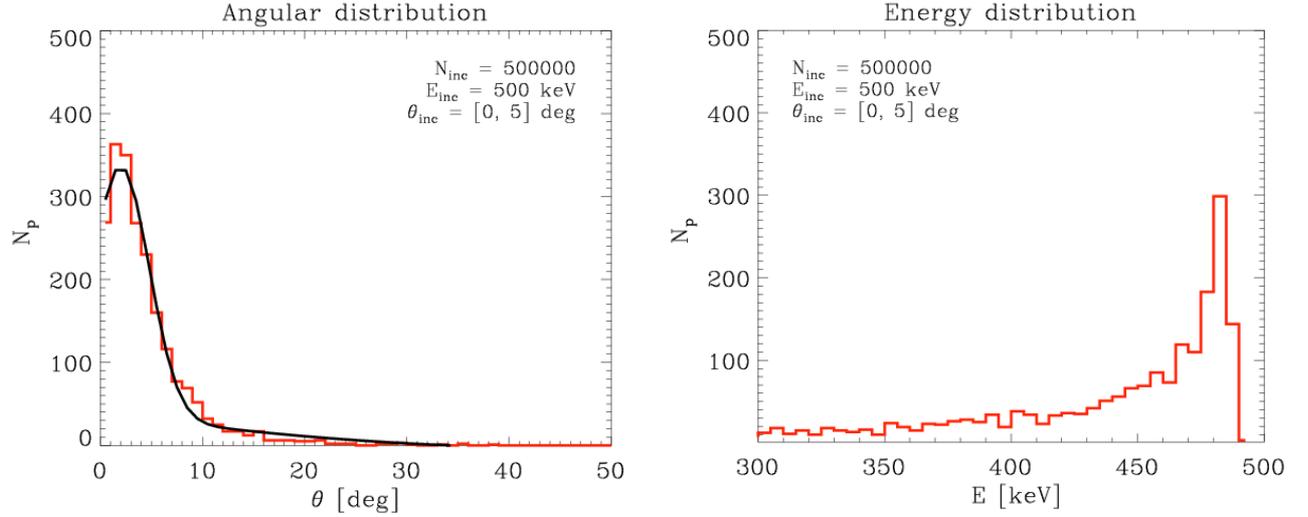

Fig. 2: some results of the application of 500 keV proton multiple scattering to SIMBOL-X mirror models, using GEANT4 libraries.
(left) The distribution of protons at the optical module exit is a Gaussian with a σ =2.8 deg.
(right) the proton energy spectrum after the interaction. The proton beam has undergone relevant energy losses after transferring the recoil momentum to the nuclei of the reflective coating.

*2.2. Firsov scattering* **of protons**

According to Dichter and Woolf[2], Lei et al.[16], the multiple scattering treatment appeared to be inefficient to explain the observed fluence of protons on the XMM detectors. Therefore, a different physical process, suggested by Firsov, was implemented in GEANT4 libraries. This process is the reflection of protons in the electron plasma outside the mirror surface. The angular distribution of reflected protons obtained by Firsov[17] is

$$N(\vartheta_i, \vartheta_s) = \frac{3}{2\pi\vartheta_i} \frac{(\vartheta_i \vartheta_s)^{3/2}}{\vartheta_s^3 + \vartheta_i^3},$$

(1)

where $\theta_i$ and $\theta_s$ are the incidence and scattering angle, respectively. The angular distribution described by the Eq. 1 is more peaked than that obtained from the multiple scattering process. Moreover, it is almost independent of the proton energy and the reflective material. In fact, it acts more like a strictly-speaking "reflection" process than a "scattering" one. Also this process has been implemented in GEANT and has been used[16] to re-calculate the proton reflection efficiency of the XMM mirrors. The Firsov scattering appears to be more efficient than multiple scattering by up to an order of magnitude.

**2.3 Proton reflection at the layer potential step**

This quantum-mechanical model treats the proton reflection as the reflection of a wave on the internal potential of the reflective coating. In fact, the internal environment of a metal (e.g. the reflective coating) is know to be at positive potential (around 15 V for many metals), as proven by the fact that, *normally,* conduction electrons do not leave spontaneously the metal. Aschenbach[10] relates the inner potential, $V_0$, to the refractive index of particles with charge $q$ and kinetic energy $K$,

$$n = \sqrt{1 - q\frac{V_0}{K}},$$

(2)

and since for protons $q > 0$, $n < 1$, if $\cos \theta_i > n$, i.e. $\theta_i < (qV_0/K)^{1/2}$, they undergo a total reflection. This can also be seen from a different viewpoint. For *electrons*, whose charge is negative, the potential is attractive. According to quantum mechanics, they have a likelihood to be reflected at the potential step, given by

$$R = \left(\frac{k_{\perp 2} - k_{\perp 1}}{k_{\perp 2} + k_{\perp 1}}\right)^2, \quad (3)$$

where $k_\perp = \hbar p_\perp$ is the wave number due to normal component $p_\perp$ of the particle, in the reflective coating and in the vacuum. Denoting with $K_\perp = p_\perp^2/2m$ the kinetic energy associated to $p_\perp$, this can be also written as

$$R = \left|\frac{1 - \sqrt{1 + eV_0/K_\perp}}{1 + \sqrt{1 + eV_0/K_\perp}}\right|^2, \quad (4)$$

which is always < 1, hence electrons cannot be totally reflected. For *protons*, the step is repulsive, therefore the reflection coefficient becomes

$$R = \left|\frac{1 - \sqrt{1 - eV_0/K_\perp}}{1 + \sqrt{1 - eV_0/K_\perp}}\right|^2. \quad (5)$$

This equals 1 for $K_\perp < eV_0$, i.e., $\theta_i < (qV_0/K)^{1/2}$. Therefore, this mechanism yields a very efficient reflection without either angular dispersion or energy losses. The non-zero angular dispersion observed in proton reflectivity experiments can, indeed, be ascribed to proton scattering from mirror surface roughness, in a manner completely analogous to X-ray photon scattering.

## 3. A POSSIBLE DESIGN FOR THE SIMBOL-X PROTON DIVERTER

We want now to evaluate the feasibility of a magnetic proton diverter for SIMBOL-X. With respect to the electron diverter of Newton-XMM, there are some differences:

1) The required angular deflection, $\theta$, to a particle with mass $m$ and charge $q$ is related to the intensity, $B$, and depth, $l$, of the magnetic field through the formula[14]

$$Bl = \frac{\sqrt{2mK}}{q} \tan \vartheta. \quad (6)$$

   As real magnetic fields are seldom uniform, this means then the integrated magnetic field seen by a proton along its trajectory has to be much more intense than for electrons, since protons are 1836 times heavier. This can be obtained with higher intensity, $B$, or with a larger thickness, $l$, of the field.

2) The minimum deflection required to avoid a proton to hit the LED (8 cm - sized, 20 m far from the optics) is 0.12 deg, if one assumes that the beam is perfectly collimated and purely deflected outwards, to be compared with the minimum deflection angle for the PN-EPIC (6 cm-sized, 7.5 m from the optics), 0.23 deg. Also the reflection angles of SIMBOL-X are smaller by a factor 2 than those of XMM. However, the angular dispersion due to proton scattering has to be added to these numbers. For the XEUS X-ray telescope Turner[14] assumes an angular spread with a $\sigma = 0.6$ deg (which actually dominates the other terms) and derives as a requirement for the product $Bl = 1700$ G·cm, more or less 3 times as large as that of Newton-XMM. A similar requirement may be assumed for SIMBOL-X as long as the proton scattering is assumed to be so relevant with respect to the other terms.

3) The optical module of SIMBOL-X will have 24 spider spokes, therefore 24 magnets can be aligned on a support – that matches the spider structure - to yield the azimuthally-directed magnetic field, instead of 12 like in the case of Swift-XRT and Newton-XMM. This doubles the magnetic field available, if magnets with the same magnetization and thickness as those of Newton-XMM are used. A further freedom degree is given by the height of magnets, on which we can rely to increase the field depth, $l$. The spider spokes, indeed, will be thinner than those of the XMM optical modules and with a variable thickness (3 to 6 mm going from the axis outwards). In order to avoid a X-ray flux obstruction, the magnets cannot be wider than the spokes themselves. This implies that the magnets should either have a variable thickness (seldom available commercially) or be segmented into smaller pieces, with different thickness, aligned along each spoke (see Fig. 3).

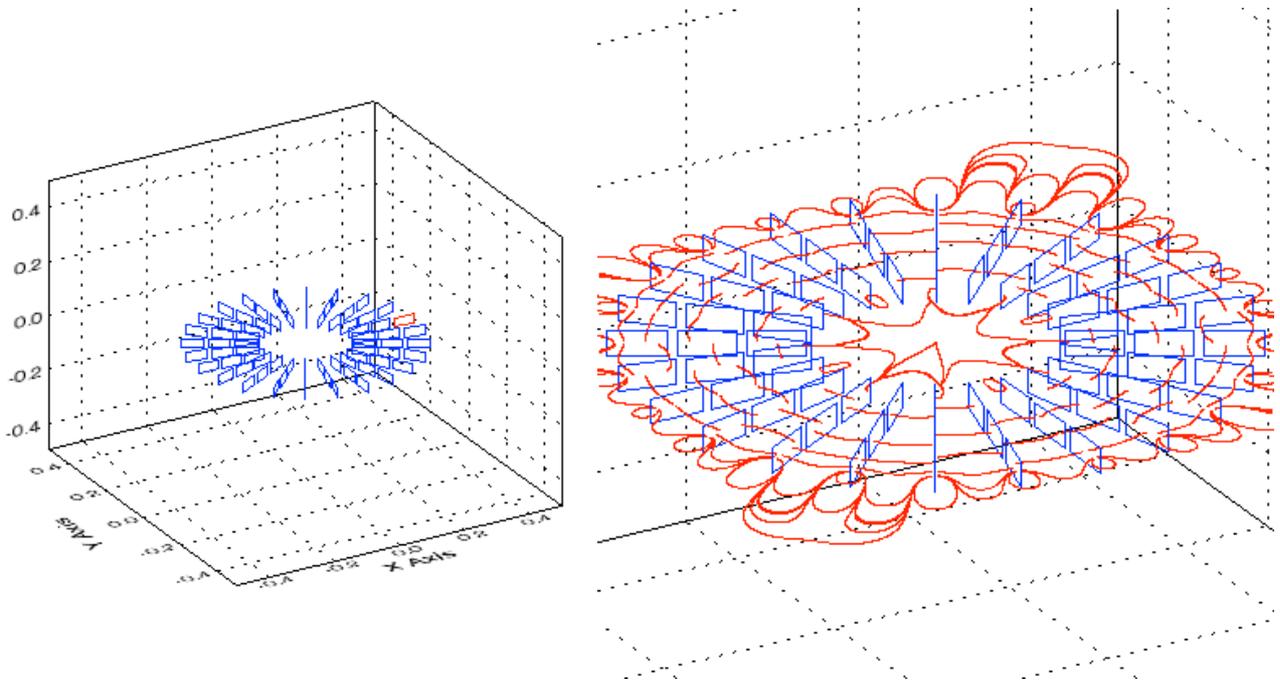

Fig. 3: (left) arrangement of the 3×24 rectangular magnetic bars for a possible implementation of the SIMBOL-X magnetic diverter. (right) some magnetic field lines traced in the diverted mid-plane. Between the magnets, the magnetic field is essentially oriented in the azimuthal direction.

There are some general requirements to the device to be designed:
1) the magnetic field should effectively prevent protons and electrons from reaching the LED in its sensitivity band. The maximum energy of protons to deflect is ~ 25 keV, unless the magnetic diverter is placed before the filters/thermal blankets[18]. In this case, the energy of a proton interacting with the LED will be higher at the interaction with the magnetic field and its deflection will be smaller. A $10^{-3}$ attenuation can be assumed as a first requirement, with a $10^{-6}$ goal[14].
2) the magnetic field should not be so extended and intense to magnetize and deform ferromagnetic elements of the optical module, like the mirror shells (Nickel) and the spider (stainless steel) ;
3) the magnetic field should not be so extended to affect the proper working the telemetric/control devices onboard the two SIMBOL-X spacecrafts;
4) the system of magnets has to be stable and should not induce stress or deformations of the optical module, owing to their mutual attraction/repulsion;
5) the magnetic diverter should fulfill the mass requirements of the mission.

A possible magnetic configuration is based on rectangular magnetic bars aligned on a radial structure in a with the same size of the mirrors spider. The supporting structure should be non-ferromagnetic to avoid the short-circuit of the magnetic flux through it. Along each radius, three bars are aligned with centers at 16, 23, 30 cm from the optical axis: a 3D representation of the bars in the box used for simulation is provided in Fig. 3. The magnets are 6 cm long and

3.5 cm high, and 3, 4, 5 mm thick for the inner, intermediate and outer ring respectively. The three magnets are separated by 1 cm in order to leave room for their fixtures. Following the Newton-XMM and Swift design[10], their magnetic moments are along the smaller side and lie in the plane of the diverter to generate a nearly-azimuthal magnetic field (Fig. 3, right). Since the particles velocity is essentially directed axially, the magnetic force is essentially radial and directed outwards for protons (inwards for electrons).

The magnets are assumed to be characterized by an uniform magnetization M. In order to generate an intense field, hard magnetic materials, like *Samarium-Cobalt magnets* (SmCo$_5$) or *Neodymium-Iron-Boron* magnets (Nd$_3$Fe$_{14}$B) should be adopted. Commercial, sintered Neodymium magnets, in particular, exhibit the following properties[19]:

a) *the highest residual magnetism*: magnetic grade N40 Neodymium magnets can reach a remanence as high as $M = 1.29$ T $= 1.03 \times 10^6$ A/m;
b) *very high coercive strengths* $H_c$, up to -1000 kA/m: this means that the magnetization is only a little affected by external magnetic fields (like the one generated by the other magnets of the diverter) within large margins;
c) they are stable in time and can operate up to temperatures as high as +80 °C.

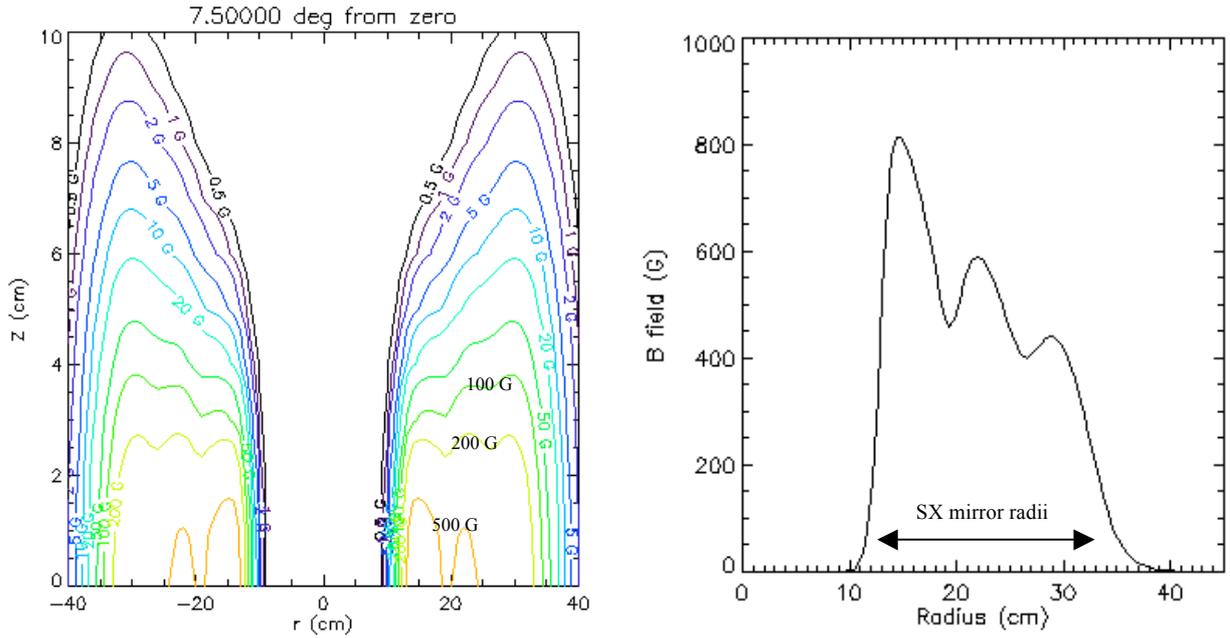

Fig. 4: (left) contour plots of the magnetic field intensity along a median plane of a sector from the diverter mid-plane out to 10 cm along the axis. The magnetic field is very intense (> 500 G) within 3 cm around the mid-plane, then it decays rapidly. At a 10 cm distance, it becomes comparable to the geomagnetic field.
(right) The profile of the magnetic field intensity in the mid-plane, along a median sector, where the magnetic field has the minimum intensity. The radial extension of the field is sufficient to cover the radii of the SIMBOL-X mirrors. Note the three B intensity peaks, related to the positions of the three magnets along the spokes.

On the other side, *Samarium-Cobalt* magnets exhibit even higher intrinsic coercitivities and can operate also at higher temperatures (even +250 °C), but they are usually characterized by weaker magnetic remanences. Moreover, they are usually more expensive and fragile, and stability against high temperatures is not usually an issue since the magnets will operate at low temperatures (around -70 °C).

The magnetic field $\underline{H}$ generated by each magnetic bar at the generic position $\underline{r}$, outside *or* inside the magnets, can be computed analytically, as a superposition of elementary dipole fields (SI units):

$$d^3\underline{H} = \frac{1}{4\pi}\left(3\frac{\underline{M}\cdot(\underline{r}-\underline{r}')}{|\underline{r}-\underline{r}'|^5}(\underline{r}-\underline{r}') - \frac{\underline{M}}{|\underline{r}-\underline{r}'|^3}\right)d^3\underline{r}'. \tag{7}$$

Even if the integral can be computed, the resulting expressions for the field are too lengthy to reproduce here. Then, the magnetic field generated by all the magnets assembly is obtained by summing the contributions of all magnetic bars. The vector $\underline{B}$ is then computed as $\mu_0\underline{H}$ in vacuum and $\mu_0(\underline{H} + \underline{M})$ inside the magnets. As expected, the resulting field $\underline{B}$ is continuous at the magnets interface, the fields $\underline{B}$ and $\underline{H}$ are parallel in vacuum and antiparallel in the magnetic material.

Assuming now for M a value slightly smaller than the magnetic remanence (1.24 T) to account for self-demagnetization, we can compute that the $B$ intensity, close to the magnets in the diverter mid-plane, is quite high: 1290 to 1400 G moving from the inner to the outer ring. Due to the symmetric geometry and the continuity of the normal component of $\underline{B}$, this is also the intensity of $B_m$ *inside* the magnets. The field $\underline{H}$ inside the magnets can be also derived from the Ampere's law (in absence of electric currents),

$$\oint_C \underline{H} \cdot \underline{dl} = 0, \qquad (8)$$

where C is chosen as a circular loop with radius $r$ in the diverter mid-plane, passing through all magnets (with thickness $\tau$) of a ring of the diverter. If one assumes that $\underline{H}$ and $\underline{B}$ keep fairly constant when crossing the magnet thickness $\tau$, one derives from Eq. 8 the $H$ value in the magnets at the radius $r$,

$$H_m = -\frac{1}{24\tau} \int_{C-outer} \underline{H} \cdot \underline{dl}, \qquad (9)$$

where the subscript "outer" means that the integral in Eq. 9 is to be restricted to the parts of the loop outside the magnets. We then obtain $H_m$ values -880 to -870 kA/m, going from the inner to the outer ring. The obtained values of $H_m$ and $B_m$ fall quite exactly on the hysteresis cycle of commercial N40-grade Neodymium magnets at 20 °C. Moreover, for the obtained value of $H_m$ the corresponding magnetization nearly equals 1.24 T, then we conclude that the assumption on the material magnetization should be a reasonable one.

In the diverter mid-plane, $B$ is maximum close to the magnets surface and minimum along the median line of the sectors. The intensity of magnetic field in the median plane of sectors is depicted in Fig. 4. Note the rapid decay of $B$ from 800 G to 0.5 G at 10 cm from the diverter plane. Since this is essentially the intensity of the geomagnetic field, that is known to not affect the shape of Nickel X-ray optics, it is sufficient to keep the optics at a 10 cm distance from the diverter to avoid mirror shell deformation due to Nickel magnetization.. The magnetic field also decreases quickly for large and small radii (and in particular it vanishes along the optical axis), so we can say that the intense B field is chiefly confined in the space between magnets.

The total mass of magnets is approximately 4.5 kg. The magnetic forces acting on magnets should also be considered. However, due to the symmetry of the system, most forces between the magnets tend to cancel out, improving the assembly stability. One can easily calculate that the residual forces acting on magnets are purely *radial* with intensity -4.0 N, -3 N, +4.9 N for the magnets of the inner, intermediate, outer ring respectively (the minus sign denotes a force directed inwards). Concerning the effectiveness of the magnetic field, the integral of the B field intensity along a vertical, straight line can be computed for comparison with the XEUS proton diverter requirement (see Eq. 6). The calculation for the actual magnetic configuration returned a product $Bl$ of 3000 G·cm for a particle crossing the diverter in the median plane of a sector close to the minimum mirror radius (14 cm from axis) to 2100 G·cm at 31 cm, close to the maximum mirror radius. At intermediate radii, the product is even higher. Thus the present design should be effective at blocking soft protons. This will be checked by means of a Monte-Carlo routine in the next section.

## 4. A MONTE-CARLO CODE TO SIMULATE THE MAGNETIC DIVERTER

To assess the effectiveness of the magnetic diverter described in the previous section, analytical formulae are seldom suitable, because magnetic fields that can be realistically obtained with permanent magnets are highly non-uniform and the proton trajectory is not easily computable with analytical methods.

We have thus written a simple numerical Monte-Carlo routine in IDL language. The program traces the trajectories of a large number of protons or electrons through the field generated by the magnetic diverter under consideration. The particle trajectory is computed numerically by integrating step-by-step the relativistic (to account for high energy electrons) equation of motion in an inhomogeneous magnetic field, whereas the initial conditions (position, direction, kinetic energy) are selected at random, within the constraints specified below. After launching a sufficiently large number of particles, one can evaluate the fraction that impinge on the detector area. This approach also takes the advantage of providing with a realistic picture of the telescope irradiation, if the actual spectrum of the particles (e.g., Fig. 1 - right) is adopted for the simulation.

For each particle, random initial conditions are selected: initial velocity direction, kinetic energy, initial coordinates x and y within the cross-section of the mirror module (i.e. with radii between 0.13 and 0.32 m from the optical axis). The distribution in x and y is assumed to be uniform, the energy distribution matches the desired particle spectrum (for this simulation the spectrum has been assumed to be flat). The particles originate at z = 20 cm of distance from the diverter plane, where the magnetic force exerted on them is expected to be negligible. For the angular distribution, we adopted a proton beam converging to the center of the detector with an additional Gaussian spread of the $\theta$ angle (i.e. the angle between the trajectory and the optical axis) with $\sigma$ =0.6 deg, following Turner[14].

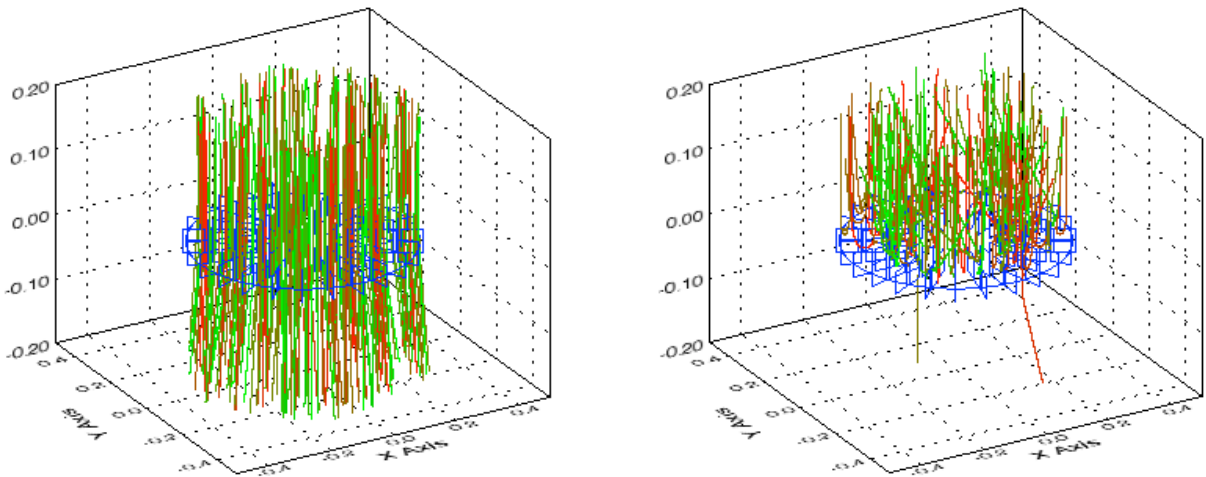

Fig. 5: (left) the simulated magnetic diverter acting on a 10 to 100 keV proton beam. Diversion of the beam is visible, although not so apparent as for the case of the (right) simulation for a 10 to 100 keV electron beam. Almost all electrons are repelled

The written IDL code has been tested by simulating the incidence of a large (1000 per run) number of particles in selected energy bands and evaluating the reduction of the number of particles falling into the detector area. This allows the evaluation of the efficiency of the diverter as a function of the particles energy. The effect of the magnetic field is illustrated in Fig. 5 for protons (left) and electrons (right). For protons the diversion is not so apparent as the deflection angles are small (even though they are large when compared to the reflection angles of mirror shells). For electrons the magnetic field has the effect of bending the trajectory so strongly to reverse their motion. This is not so strange, as for them the typical radius of curvature often becomes smaller than the magnetic field depth. Almost all electrons are repelled by the diverter: only a small fraction passes through, but it is so strongly diverging that the likelihood that they reach the detector area should be negligible.

A preliminary simulation has been performed without magnetic field to evaluate the proton beam divergence effect. The result is the proton beam divergence alone causes an attenuation of the proton flux on the detector to only 19%, insufficient to lower the proton background to acceptable values. Then we computed the impact points of protons falling into the detector area after being deviated by the magnetic field. The results are plotted in Fig. 6. The energy bands considered are 10-50 keV and 900-1100 keV. For the 10-50 keV band, the magnetic field seems to be very effective at deflecting soft protons, since none of them falls inside the detector area for the lowest energy band, which is superposed to the LED sensitivity bandwidth. This yields a probable attenuation factor of $< 10^{-3}$, which is in line with the requirement for the XEUS proton diverter[14]. The diverter becomes less effective at higher energies (see Fig. 6, lower right), as the curvature radii decrease for increasing energies. Note that these results are referred to the specific case of an angular dispersion of the reflected beam with a 0.6 deg rms. The effectiveness of the magnetic diverter is, in fact, quite sensitive to the proton beam collimation.

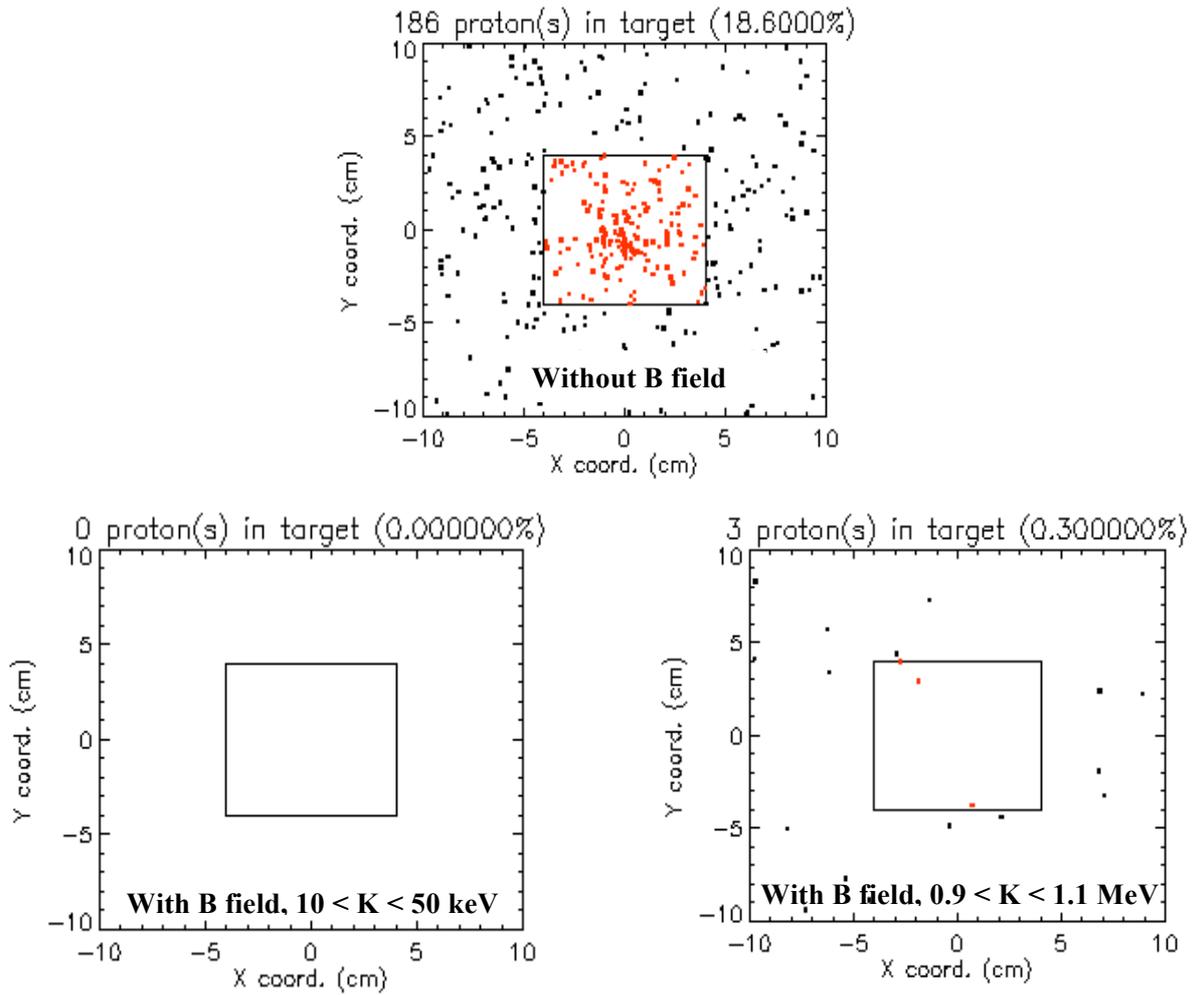

Fig. 6: results of the simulation with the magnetic diverter with 1000 protons in 2 different proton energy bands, assuming a Gaussian spread of protons with σ = 0.6 deg. For comparison we plotted also the results of the simulation without magnetic field in the upper panel. The small square represents the SIMBOL-X LED area. The effectiveness of the magnetic diverter is clearly seen.

## 5. CONCLUSIONS AND OPEN ISSUES

In this work we have proposed a possible implementation of the magnetic diverter for soft protons and electrons in order to reduce the particle background in the SIMBOL-X telescope. Like the electron diverter of Newton-XMM or Swift-XRT, it is based on commercially-available permanent magnets with a very high magnetization. We have computed the magnetic field achievable with a given arrangement of magnetic bars and we have checked its effectiveness for soft protons and electrons by means of a Monte-Carlo simulation in IDL language. The results are in reasonable agreement with previous calculations[14].

While this was a preliminary feasibility study, further developments of this work will include more factors in order to provide with a more realistic approximation to the real behavior of the equipment. Future actions will include:

1. establishing the physical process responsible for proton reflection/scattering in order to return a more realistic simulation of the angular spread of protons at the magnetic diverter entrance;

2. coupling of the GEANT and IDL codes to obtain a self-consistent picture of the proton-telescope interaction;

3. accounting for the proton spectrum modification caused by the transmission through the thermal blankets[18];
4. establishing the thermo-mechanical properties and stability of the entire magnetic diverter;
5. investigation of alternative magnetic configurations (smaller magnets, magnets scaled along the z axis, combination of magnets made of different magnetic materials, asymmetrical configurations);
6. production and test of a proton diverter prototype for SIMBOL-X for an experimental verification of the magnetic field effectiveness.

## ACKNOWLEDGMENTS

We thank G. Parodi, M. Ottolini (BCV Progetti, Milano), B. Aschenbach (MPE, Garching), S. Mereghetti, S. Molendi (IASF Milano) for useful suggestions and discussions.